\documentclass[runningheads]{llncs}

\usepackage{eccv}

\usepackage{graphicx}
\usepackage{booktabs}
\usepackage[accsupp]{axessibility}  
\usepackage{hyperref}

\usepackage{eccvabbrv}

\usepackage{amsmath,amsfonts,bm}

\def\eqref#1{equation~\ref{#1}}

\def\1{\bm{1}}

\def\vv{{\bm{v}}}

\DeclareMathAlphabet{\mathsfit}{\encodingdefault}{\sfdefault}{m}{sl}
\SetMathAlphabet{\mathsfit}{bold}{\encodingdefault}{\sfdefault}{bx}{n}

\usepackage{wrapfig}
\usepackage{picinpar}
\usepackage{color}
\usepackage{multirow}
\usepackage[flushleft]{threeparttable} 
\usepackage{boldline} 
\usepackage{algorithm}

\usepackage{balance}

\usepackage{dcolumn}
\usepackage{bm}
\usepackage[utf8]{inputenc}
\usepackage[T1]{fontenc}

\usepackage[noend]{algpseudocode}
\usepackage{colortbl}

\usepackage{tikz}
\usepackage{calc}
\usepackage{multirow}
\usepackage{multicol}
\usepackage{bbding}
\usepackage{caption}
\usepackage{mathtools}

\usepackage{newfloat}
\usepackage{listings}

\usepackage{orcidlink}

\newcommand{\Bmat}{{\bf B}}

\newcommand{\Dmat}{{\bf D}}

\newcommand{\Kmat}[0]{{{\bf K}}}

\newcommand{\Mmat}[0]{{{\bf M}}}

\newcommand{\Pmat}[0]{{{\bf P}}}
\newcommand{\Qmat}[0]{{{\bf Q}}}

\newcommand{\Umat}{{{\bf U}}}

\newcommand{\Vmat}[0]{{{\bf V}}}
\newcommand{\Wmat}[0]{{{\bf W}}}
\newcommand{\Xmat}{{\bf X}}
\newcommand{\Ymat}[0]{{{\bf Y}}}
\newcommand{\Zmat}{{\bf Z}}

\newcommand{\Av}{\boldsymbol{A}}

\newcommand{\bv}{\boldsymbol{b}}
\newcommand{\cv}{{\boldsymbol{c}}}

\newcommand{\Iv}[0]{{\boldsymbol{I}}}

\newcommand{\xv}{\boldsymbol{x}}
\newcommand{\yv}{\boldsymbol{y}}

\newcommand{\zv}{\boldsymbol{z}}

\newcommand\nnfootnote[1]{%
  \begin{NoHyper}
  \renewcommand\thefootnote{}\footnote{#1}%
  \addtocounter{footnote}{-1}%
  \end{NoHyper}
}

\newcommand{\inv}{^{-1}}
\def\eg{e.g}

\definecolor{color3}{rgb}{0.898,0.898,0.902}

\definecolor{color2}{rgb}{0.85,0.85,0.90}
\definecolor{color1}{rgb}{0.85,0.95,0.6}

\newcommand{\done}[1]{\textcolor{green}{\checkmark}}

\begin{document}

\title{Latent Diffusion Prior Enhanced Deep Unfolding for Snapshot Spectral Compressive Imaging}

\titlerunning{Latent Diffusion Prior Enhanced Deep Unfolding}

\author{Zongliang Wu\inst{1,2*} \orcidlink{0000-0003-0750-0246} \and
Ruiying Lu\inst{3*}\orcidlink{0000-0002-8825-6064} \and
Ying Fu\inst{4}\orcidlink{0000-0002-6677-694X}\and
Xin Yuan\inst{2}\textsuperscript{(\Envelope)}\orcidlink{0000-0002-8311-7524}
}

\authorrunning{Z.~Wu et al.}

\institute{Zhejiang University, Hangzhou, China \and
School of Engineering, Westlake University, Hangzhou, China.
\email{\{wuzongliang,xyuan\}@westlake.edu.cn}\\
\and
School of Cyber Engineering, Xidian University, Xi'an, China. 
\email{ruiyinglu\_xidian@163.com}  
\and
School of Computer Science and Technology, Beijing Institute of Technology, Beijing, China.  \email{fuying@bit.edu.cn}
}

\maketitle
\nnfootnote{*Z. Wu and R. Lu—Contribute equally}

\begin{abstract}
Snapshot compressive spectral imaging reconstruction aims to reconstruct three-dimensional spatial-spectral images from a single-shot two-dimensional compressed measurement.
Existing state-of-the-art methods are mostly based on deep unfolding structures but have intrinsic performance bottlenecks: $i$) the ill-posed problem of dealing with heavily degraded measurement, and $ii$) the regression loss-based reconstruction models being prone to recover images with few details. In this paper, we introduce a generative model, namely the latent diffusion model (LDM), to generate degradation-free prior to enhance the regression-based deep unfolding method by a two-stage training procedure. Furthermore, we propose a Trident Transformer (TT), which extracts correlations among prior knowledge, spatial, and spectral features, to integrate knowledge priors in deep unfolding denoiser, and guide the reconstruction for compensating high-quality spectral signal details.
To our knowledge, this is the first approach to integrate physics-driven deep unfolding with generative LDM in the context of CASSI reconstruction. Comparisons between synthetic and real-world datasets illustrate the superiority of our proposed method in both reconstruction quality and computational efficiency. The code is available at \url{https://github.com/Zongliang-Wu/LADE-DUN}.
\end{abstract}
\section{Introduction}
In contrast to normal RGB images which only have three spectral bands, hyperspectral images (HSIs) contain multiple spectral bands with more diverse spectral information. The spectral information serves to characterize distinct objects assisting high-level image tasks~\cite{van2010tracking,uzkent2017aerial,li2019deep,li2020prior,rao2022siamese} and the observation of the world like medical imaging~\cite{lu2014medical,ul2021review} and remote sensing~\cite{goetz1985imaging,lu2020rafnet}. However, the capture of HSIs is a question that has been studied for a long time because we need to collect 3-dimensional (3D) HSI signals by 2D sensors.
\begin{figure}[t] 
\centering
\begin{minipage}[t]{0.49\textwidth}
\includegraphics[width=1\textwidth]{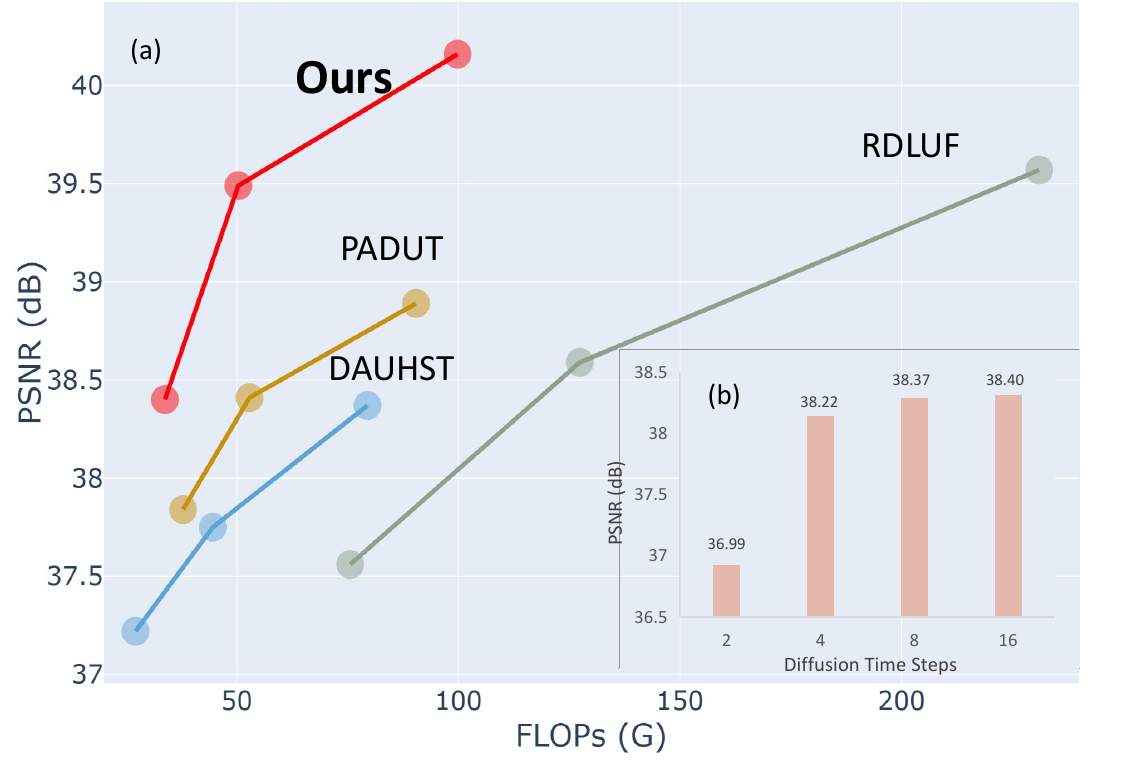}
\caption{\small (a) Comparison of PSNR (dB)-FLOPs (G) with previous HSI reconstruction methods. (b) The ablation study of using different time steps in diffusion. Our method achieves the desired results by only very few steps.}
\label{fig:teaser_bubble}
\end{minipage}
\begin{minipage}[t]{0.49\textwidth}
\centering
\includegraphics[width=1\textwidth]{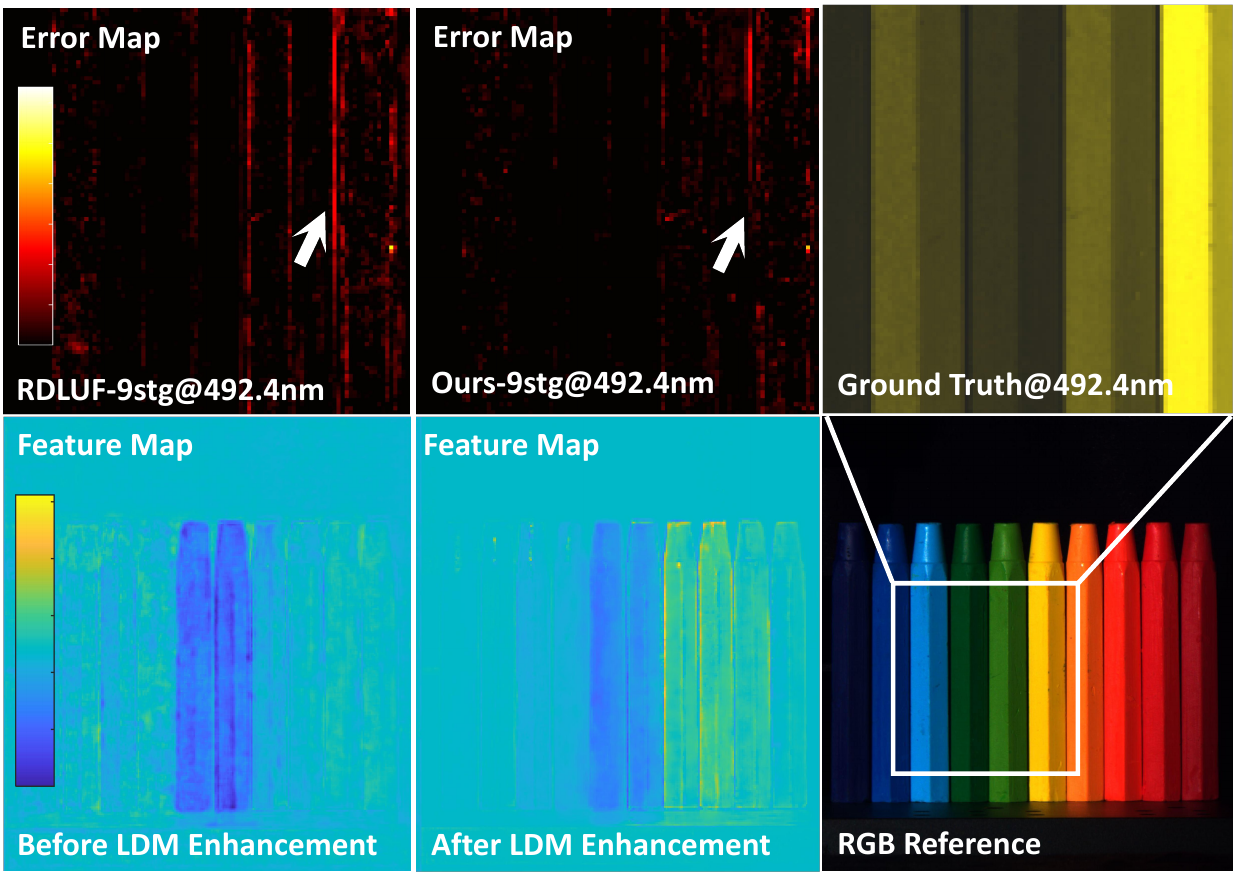}
\caption{\small The top row: the error maps of the previous SOTA and our method. The bottom row is the feature map before and after applying LDM enhancement. The enhanced features demonstrate less noise and clearer edges.}
\label{fig:teaser_vis}
\end{minipage}
\end{figure}

For many years, scientists have focused on how to collect HSIs in a quick and convenient method. In 2007, based on compressive sensing theory, a single-shot compressive spectral imaging way~\cite{Gehm07} was created to efficiently collect HSIs, named coded aperture snapshot spectral imaging (CASSI). The later improvement works~\cite{wagadarikar2008single,meng2020end} provide better imaging quality and lower cost.
CASSI modulates the HSI signal across various bands and combine all the modulated spectra to produce a 2D compressed measurement. Consequently, the task of reconstructing the 3D HSI signals from the 2D compressive measurements presents a fundamental challenge for the CASSI system.

The reconstruction process can be viewed as solving an ill-posed problem. Many attempts at solving this problem including traditional model-based methods~\cite{Bioucas-Dias2007TwIST,Beck09IST,Yuan16ICIP_GAP} and the learning-based methods\cite{charles2011learning,Miao19ICCV,Meng20ECCV_TSAnet} have been proposed since the inception of CASSI system.
The deep unfolding network (DUN) is a combination of convex optimization and neural network prior (denoiser), enjoying both the interpretability of the model-based method and the power of learning-based methods. This branch of methods leads the development trend in recent years~\cite{wang2022snapshot,cai2022degradation,meng2023deep,dong2023residual,li2023pixel} and achieves SOTA performance.

However, unlike super-resolution or deblurring that recovers from natural images, CASSI reconstruction has to recover HSIs from the compressed domain measurements, which results in severe degradation according to physical modulation, spectral compression, and multiple types of system noises. Thus, the CASSI reconstruction problem is much harder to learn intrinsic HSI properties than the normal image restoration tasks~\cite{wang2020fusionnet,zhang2021hyperspectral,lu2022heterogeneity,zhang2022joint,li2023spatial,li2023spectral,lai2024hyperspectral}. In the unfolding framework of the CASSI reconstruction method, the denoising network plays a critical role in deciding the final performance, which is embedded in each stage of the DUN. However, it always suffers from the performance bottleneck due to the intrinsic ill-posed problem of dealing with heavily degraded measurements.
Thus, a high-performance denoiser with degradation-free knowledge is desired for CASSI reconstruction.
Another problem is that previous popular regression-based reconstruction methods have difficulty in recovering details because the widely used
regression losses are conservative with high-frequency details~\cite{saharia2022image}.

To address these challenges, we introduce a generative prior in this paper to guide the reconstruction process in an unfolding framework. During training, the prior will be first learned from clean HSIs by an image encoder and then generated by a Latent Diffusion Model (LDM) from Gaussian noise and compressed measurement. Then, the learned prior is embedded into the deep denoiser of the DUN by a prior-guided Transformer. 
Significantly, our DUN is able to leverage external prior knowledge from clean HSIs and the powerful generative ability of LDM enhancing its reconstruction performance. The primary contributions presented in this paper can be summarized as follows:
\begin{itemize}
\setlength{\itemsep}{0pt}
    \item [$i$)] We propose a novel {\bf LDM-based unfolding network} for CASSI reconstruction, where the clean image priors are generated by a latent diffusion model to facilitate high-quality hyperspectral reconstruction. No additional data or training time is required. To the best of our knowledge, this is the first attempt to combine the physics-driven deep unfolding with generative LDM in CASSI reconstruction.
    \item [$ii$)] We design a three-in-one Transformer structure dubbed Trident Transformer (TT) to extract the correlation among prior knowledge, spatial, and spectral features. In TT, motivated by pansharpening techniques, we introduce an asymmetric cross-scale multi-head self-attention (ACS-MHSA) mechanism designed to efficiently fuse spatial-spectral features.
    \item [$iii$)] Extensive experiments on the synthetic benchmark and real dataset demonstrate the superior quantitative performances (Fig.~\ref{fig:teaser_bubble}), visual quality (Fig.~\ref{fig:teaser_vis}), and lower computational cost of our proposed method.
\end{itemize}
\section{Related Work}
\subsection{Diffusion Model in Low-level Vision} 
Diffusion models (DMs)~\cite{ho2020denoising,sohl2015deep} are probabilistic generative models, which model the data distribution by learning a gradual iterative denoising process from the Gaussian distribution to the data distribution. Notably, they demonstrate promising capabilities in generating high-quality samples that encompass a wide range of modes, including super-resolution~\cite{gao2023implicit} and inpainting~\cite{lugmayr2022repaint}. 
In light of the impressive achievements of diffusion models in image domains, numerous research endeavors~\cite{harvey2022flexible,ho2022imagen,blattmann2023align,hoppe2022diffusion} have extended it to video generation. 
However, diffusion models suffer from significant computation inefficiency regarding data sampling, primarily due to the iterative denoising process required for inference. 
To address this challenge, several methods propose effective sampling techniques from trained diffusion models~\cite{song2020denoising,zhang2022fast}, or alternatively learning the data distribution from a low-dimensional latent space~\cite{rombach2022high}, \ie the latent diffusion model. The latent diffusion has a relatively faster speed and powerful generative ability for super-resolution and inpainting, but similar to the normal diffusion model, it is also prone to issues such as misaligned distribution of fine details and the occurrence of unwanted artifacts, leading to suboptimal performance in distortion-based metrics, e.g., PSNR. Moreover, latent diffusion costs large computational resources both for training and inference due to its large-size encoder and denoiser.
Towards this end, some works combine the generative diffusion model with the regression restoration network and work well on distortion-based metrics like deblurring~\cite{ren2023multiscale}. The recent works~\cite{chen2023hierarchical,xia2023diffir} employ LDM on many low-level vision tasks and achieve SOTA with reasonable computational cost. We name these methods `integrated diffusion' to distinguish them from the pure diffusion method. Nevertheless, employing diffusion models for the efficient reconstruction of hyperspectral images from highly compressed and degraded measurements presents significant challenges.
\subsection{Hyperspectral Image Reconstruction}
Before the advent of the deep learning wave, traditional model-based methods iteratively solved this inverse problem by convex optimization~\cite{Wang17PAMI,zhang2019computational,Wagadarikar08CASSI} with some hand-crafted constraints based on image priors, like sparsity~\cite{Kittle10CASSI} and low-rank~\cite{Liu18TPAMI}. These methods are robust and interpretable but require manual parameter tuning with low reconstruction speed and performance. With the help of deep learning, Plug-and-play (PnP) algorithms~\cite{Chan2017PlugandPlayAF, PnP2019ICML, Yuan20PnPSCI, yuan2021pnp,qiu2021effective,chen2023combining,chen2024hyperspectral}, embeds pre-trained denoising networks into convex optimization to solve the reconstruction problem, but still has limitations on performance because of the pre-trained denoiser. In recent years, the End-to-end (E2E) reconstruction directly trains a powerful deep neural network, like convolutional neural network (CNN)~\cite{cheng2022recurrent,lu2020rafnet,hu2022hdnet} and Transformers~\cite{cai2022mask,cai2022coarse,cai2024binarized}, to learn the recovery process from inputs (measurements) to outputs (desired HSIs). 
However, this simple design lacks interpretability and robustness for various hardware systems.
Therefore, an interpretable design of a reconstruction network that unfolds a convex optimization process named DUN is proposed to leverage these problems. A series of CASSI reconstruction works based on DUN~\cite{meng2020gap,ma2019deep,cai2022degradation,dong2023residual,xu2023degradation,li2023pixel} are proposed and become the state-of-the-art (SOTA) method. DUN can combine both interpretability in model-based methods and performance in deep learning-based methods to reconstruct CASSI at a fast speed.
It changes iterative steps in optimization into several stages in a single network. The prior for optimization becomes a deep neural network denoiser. Since the DUN needs to define the forward model of imaging, it is also considered a physics-driven network.
However, the recent DUNs still have bottlenecks for their regression-based denoiser design and the difficulty of dealing with compressive measurement features. 
Bearing these concerns, we propose an `integrated diffusion' module and integrate it into the physics-driven DUN framework and design an efficient way to aggregate complex features during reconstruction.
\begin{figure*}[t!]
\centering
\includegraphics[width=1\linewidth]{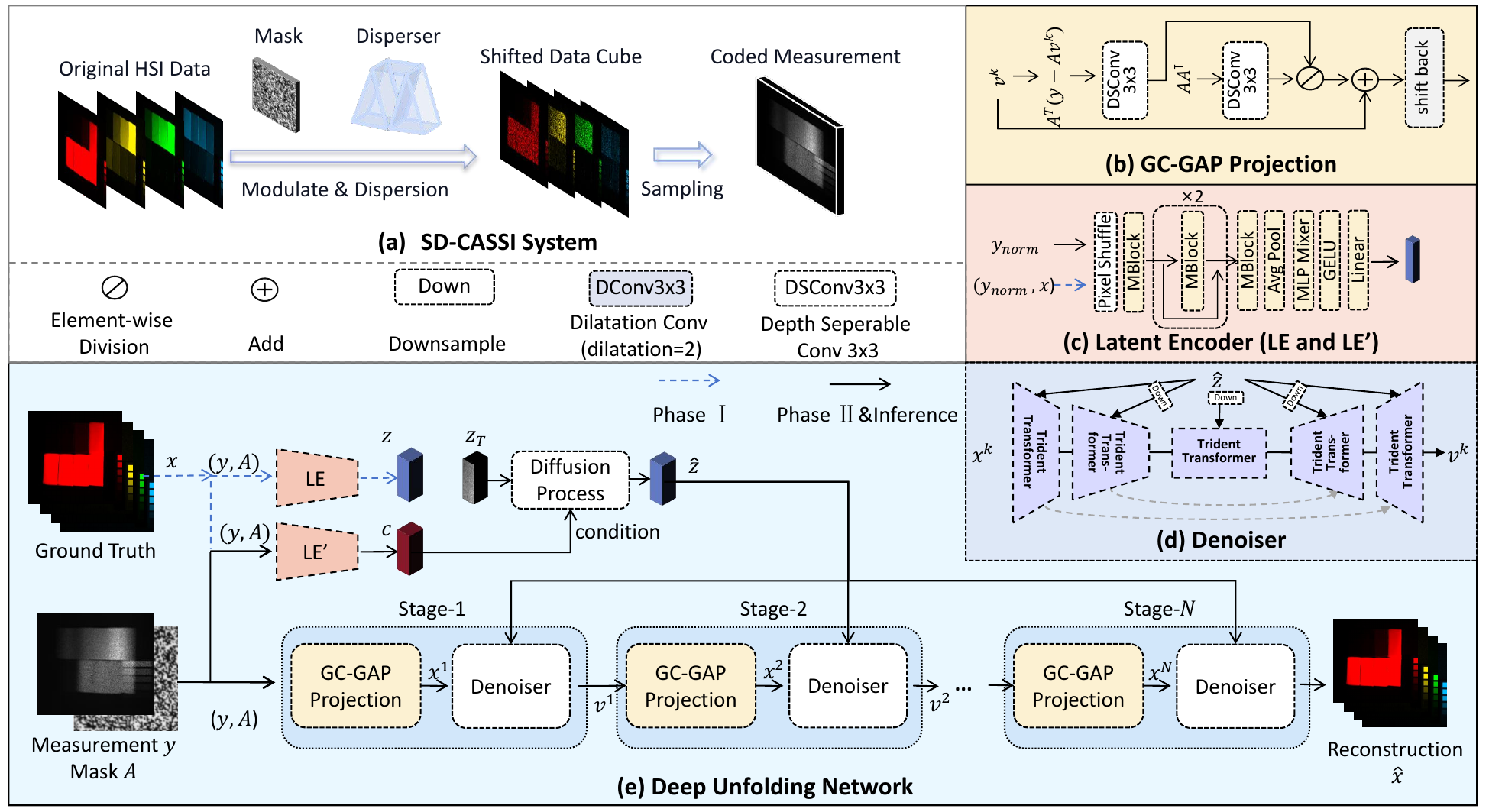}
\caption{\small (a) The single disperser CASSI imaging process. HSI data cube is captured by a monochromatic sensor. (b) GC-GAP projection. (c) Latent encoder. (d) Simplified Denoiser. (e) The measurement $\yv$ and masks $\Av$ pass through an N-stage DUN, where each stage is composed of a GC-GAP projection and a denoiser. The denoiser follows a U-shape structure and consists of five Trident Transformers (TT), where each TT is assisted with prior knowledge $\zv_{GT}$ generated from the diffusion model.}

\label{fig:overall}
\end{figure*}
\section{Problem Formulation}
The CASSI system has high efficiency in capturing 3D spectral signals by initially coding spectral data with different wavelengths in an aperture and then integrating them into a 2D monochromatic sensor.
The mathematical forward process of the widely used single-disperser CASSI (SD-CASSI)~\cite{Wagadarikar08CASSI} can be illustrated as Fig.~\ref{fig:overall} (a).
As can be seen, the original HSI data, denoted as $\bm{X}\in\mathbb{R}^{W\times H\times N_\lambda}$, is coded by the physical mask $\bm{M}\in\mathbb{R}^{W\times H}$, where $W$, $H$, and $N_\lambda$ denote the width, height, and the number of spectral channels, respectively. The coded HSI data cube is represented as $\Xmat'(:,:,n_{\lambda})=\Xmat(:,:,n_{\lambda})\odot \Mmat, n_{\lambda}=1,2,\ldots,N_\lambda$,
where $\odot$ represents the element-wise multiplication. After light propagating through the disperser, each channel of $\Xmat'$ is shifted along the $\bm{H}$-${axis}$. The shifted data cube is denoted as $\Xmat''\in\mathbb{R}^{W\times \tilde{H}\times N_\lambda}$, where $\tilde{H}=H+d_{\lambda}$. $d_{\lambda}$ is the shifted distance of the $N_\lambda$-th wavelength.
This process can be formulated as modulating the shifted version $\tilde{\Xmat}\in\mathbb{R}^{W\times \tilde{H}\times N_\lambda}$ with a shifted mask $\tilde{\Mmat}\in\mathbb{R}^{W\times \tilde{H}\times N_\lambda}$, where $\bm{\tilde{M}}(i,j,n_{\lambda})=\bm{M}(w,h+d_{\lambda})$.
At last, the imaging sensor captures the shifted image into a 2D measurement $\Ymat$, calculated as
$
\Ymat=\textstyle \sum _{n_{\lambda}=1}^{N_\lambda} {\tilde{\Xmat}}(:,:,n_{\lambda})\odot {\tilde{\Mmat}}(:,:,n_{\lambda})+\Bmat,
$
where $\Bmat\in\mathbb{R}^{W\times \tilde{H}}$ denotes the measurement noise. By vectorizing the data cube and measurement, that is $\xv=\mathtt{vec}(\tilde{\Xmat})\in\mathbb{R}^\mathit{W\tilde{H}N_\lambda}$ and $\yv=\mathtt{vec}({\Ymat})\in\mathbb{R}^\mathit{W\tilde{H}}$, this model can be formulated as
\begin{equation}
  \yv=\boldsymbol{A} \xv+ \bv, \label{eq:linear}
\end{equation}
where $\boldsymbol{A}\in\mathbb{R}^{\mathit{W\tilde{H}}\times \mathit{W\tilde{H}N_\lambda}}$ denotes the sensing matrix (coded aperture) which is a concatenation of diagonal matrices, that is
    $\boldsymbol{A} = [\Dmat_1,\dots,\Dmat_\lambda]$, 
where $\Dmat_\lambda= Diag(vec(\tilde{\Mmat}(:, :, n_{\lambda})))$ is the diagonal matrix with $vec(\tilde{\Mmat}(:, :, n_{\lambda}))$ as the diagonal elements. 
In this paper, we will propose a method to solve 
the ill-posed problem, reconstructing the HSI $\xv$ from the compressed measurement $\yv$.
\section{Proposed Model}
To solve the problem in Eq.~(\ref{eq:linear}), we proposed a novel unfolding enhanced by latent diffusion prior. As shown in Fig.~\ref{fig:overall}(e), in the inference phase, the measurement and masks pass through an N-stage DUN, where each stage is composed of a GC-GAP projection and a denoiser. The denoiser follows a U-shape structure and consists of Trident Transformers (Fig.~\ref{fig:overall}(d)), where each TT is assisted with the LDM prior. We'll describe these modules in more detail in this section.
\subsection{The Unfolding GAP Framework}
Eq.~(\ref{eq:linear}) can be typically solved by convex optimization by the objective below:
\begin{equation}
\label{object}
\hat{\boldsymbol{x}}=\textstyle \underset{\boldsymbol{x}}{\arg \min } \frac{1}{2}\|\boldsymbol{y}-\boldsymbol{A x}\|_{2}^{2}+\tau R(\boldsymbol{x}),
\end{equation}
where $\tau$ is a noise-balancing factor. The first term guarantees that the solution $\hat{\boldsymbol{x}}$ fits the measurement, and the term $R(\boldsymbol{x})$ refers to the image regularization.

To solve the optimization problem, we employ GAP (Generalized Alternating Projection) as our optimization framework, which extends classical alternating projection to the case in which projections are performed between convex sets that undergo a systematic sequence of changes. It can be interrupted anytime to return a valid solution and resumed subsequently to improve the solution~\cite{liao2014generalized}. This property is very suitable for DUN which has very limited `optimization iterations' (stages in the DUN). 
Specifically, we introduce an auxiliary parameter $\vv$, Eq.~(\ref{object}) can be written as: 
\begin{equation}
(\hat{\boldsymbol{x}},\hat{\boldsymbol{v}})=\underset{\boldsymbol{x},\boldsymbol{v}}{\arg \min } \frac{1}{2}\|\boldsymbol{x}-\boldsymbol{v}\|_{2}^{2}+\tau R(\boldsymbol{v}), \quad \text{s.t.}\quad\yv=\Av\xv.\label{object-2-gap}
\end{equation} 
Then, the problem can be solved by the following sub-problems:
Firstly, we aim at updating $\boldsymbol{x}:$ 
\begin{equation}
\boldsymbol{x}^{(k+1)}=\boldsymbol{v}^{(k)}+\boldsymbol{A}^{\top}(\boldsymbol{A} \boldsymbol{A}^{\top})^{-1}(\boldsymbol{y}-\boldsymbol{A} \boldsymbol{v}^{(k)}). \label{GAP-proj}
\end{equation}
This step projects measurement to a 3D signal space by Euclidean projection. Secondly, we aim at updating $v$: 
\begin{equation}
\boldsymbol{v}^{(k+1)}=\mathcal{D}_{k+1}(\boldsymbol{x}^{(k+1)},\zv), \label{GAP-denoise}
\end{equation}
where $\mathcal{D}_{k}$ is the neural network denoiser of the $k-th$ stage and $\zv$ is prior knowledge which will be described in Sec.~\ref{sec:DUN_LDM}. This step tries to map $\boldsymbol{x}^{(k+1)}$ to the target signal domain.
Considering the projection step Eq.~(\ref{GAP-proj}), assisted by deep network, we can modify it as follows:
\begin{equation}
\boldsymbol{x}^{(k+1)}=\boldsymbol{v}^{(k)}+\text{DSC}(\boldsymbol{A}^{\top}(\boldsymbol{A} \boldsymbol{A}^{\top})\inv(\yv - \boldsymbol{A} \boldsymbol{v}^{(k)})),  \label{GAP-GC—proj}
\end{equation}
where $\text{DSC}(\cdot)$ denotes a set of depthwise separable convolution and GELU~\cite{hendrycks2016gaussian} operations. 
The detailed process is shown in Fig.~\ref{fig:overall}(b).  
Considering the stage number is much less than the iteration numbers in traditional model-based methods, it is difficult to achieve convergence with limited steps of gradient descent. Thus, we utilize these learnable parameters to rectify the gradients in the limited stage, and we refer to this method as the Gradient Correction GAP (GC-GAP).
The overall unfolding framework is shown in Fig.~\ref{fig:overall}(e), where mask $A$ and measurement $y$ are inputs of the network. According to the Eq.~(\ref{GAP-GC—proj}) and ~(\ref{GAP-denoise}), the first stage outputs $\boldsymbol{v}^{1}$ can be obtained. 

\subsection{Latent Diffusion Prior Assisted Unfolding Denoising}\label{sec:DUN_LDM}
The denoising process in DUN leads to a natural performance bottleneck due to the intrinsic problem of heavily degraded input. Thus, we introduce external degradation-free prior knowledge to compensate for the denoising process. We will then introduce this process in a two-phase manner.

\noindent\textbf{Phase \uppercase\expandafter{\romannumeral1}: Learning Prior Knowledge from clean HSIs.} In this phase, we use an image encoder to compress both compressive measurement $\yv$ and clean HSIs (Ground-Truth hyperspectral images) $\xv$ into latent space. However, instead of simply using measurement $\yv$, we transfer $\yv$ by Euclidean projection to 3D HSIs space and normalize it by sensing matrix $\yv_{norm}\in\mathbb{R}^{W \times H \times N_\lambda} = \boldsymbol{A}^{\top}\left(\boldsymbol{A}\boldsymbol{A}^{\top}\right)^{-1}\yv$. This will improve the balance between two different inputs and easier for the encoder to learn their relation. The input of the encoder in the first phase is $\Iv_{\text{E}}\in\mathbb{R}^{W\times H\times 2N_\lambda} = \text{concatenate}(\yv_{norm},\xv)$. Thus the latent encoder process can be written as $\zv_{GT}\in\mathbb{R}^{N\times C} = \text{LE}(\Iv_{\text{E}})$, where $N\ll W\times H$, $C$ is the latent feature channel number. The $\text{LE}$ can be seen in Fig.~\ref{fig:overall} (c):  to alleviate the computational burden within LDM, we employ mobile blocks (MBlocks)~\cite{howard2019searching}, devoid of batch normalization instead of normal convolution. It aims at efficiently extracting representative visual features while maintaining computational efficiency. Moreover, considering the limitations in convolution, we add an MLP-Mixer~\cite{tolstikhin2021mlp} in LE to provide fast information exchange between patches by token-mixing MLP. Then the $\zv_{GT}$ will be used as prior in the denoiser to compensate for the denoising errors. The DUN will reconstruct HSI signals using measurement and mask with the assistance of $\zv_{GT}$, \ie $\hat{\xv} = \text{DUN}(\yv, \boldsymbol{A}, \zv_{GT}).$ Note that unlike the original LDM having its entire `Auto Encoder', there is no corresponding `decoder' specified here, because $\zv$ is sent to DUN for `decoding'. 
In this phase, we only use the reconstruction loss: 
$
\mathcal{L}_{\text {rec}} =\left\|\xv-\hat{\xv} \right\|_{1}$.

\noindent\textbf{Phase \uppercase\expandafter{\romannumeral2}: Generating Prior by Latent Diffusion Model.}
After learning the prior representation from clean HSIs, we aim to learn an LDM to generate this prior condition on measurement $\yv$ in the second phase.
Specifically, the encoder in the first phase $\text{LE}$ is fixed to encode clean HSIs and measurements to $\zv_{GT}$ as the generative object of latent space, \ie the starting point of the forward Markov process in the diffusion model. Then as usual forward process, Gaussian noise will be gradually added on $\zv_{GT}$ across $T$ time steps according to the parameter $\beta_t$:
\begin{equation}
\begin{split}
&q\left(\zv_{1: T} \mid \zv_0\right)=\textstyle \prod_{t=1}^T q\left(\zv_t \mid \zv_{t-1}\right), \forall t=1, \ldots, T,\\
&\quad q\left(\zv_t \mid \zv_{t-1}\right)=\mathcal{N}\left(\zv_t ; \sqrt{1-\beta_t} \zv_{t-1}, \beta_t \mathbf{I}\right), \label{eq:df_fwd_noise}
\end{split}
\end{equation}
where $\zv_t$ represents the noisy features at the $t$-th step, and $\zv_0=\zv_{GT}$ is the generative target. $\beta_{1: T} \in(0,1)$ are hyperparameters that control the variance of the Gaussian distribution $\mathcal{N}$. Through iterative derivation with reparameterization~\cite{Kingma2014ICLR}, Eq.~(\ref{eq:df_fwd_noise}) can be written as:
\begin{equation}
\begin{split}
&q\left(\zv_t \mid \zv_0\right)=\mathcal{N}\left(\zv_t ; \sqrt{\bar{\alpha}_t} \zv_0,\left(1-\bar{\alpha}_t\right) \mathbf{I}\right), \\
&\textstyle \quad \alpha=1-\beta_t, \quad \bar{\alpha}_t=\prod_{i=1}^t \alpha_i.
\end{split}
\end{equation}

The reverse process involves generating the prior features from a pure Gaussian distribution step-by-step condition on the measurement. 
The reverse process operates as a $T$-step Markov chain that runs backward from $\zv_T$ to $\zv_0$. Specifically, the posterior distribution of the reverse step from $\zv_t$ to $\zv_{t-1}$ can be formulated as:
\begin{equation}
\begin{split}
&q\left(\zv_{t-1} \mid \zv_t, \zv_0\right)=\textstyle \mathcal{N}\left(\zv_{t-1} ; \boldsymbol{\mu}_t\left(\zv_t, \zv_0\right), \frac{1-\bar{\alpha}_{t-1}}{1-\bar{\alpha}_t} \beta_t \mathbf{I}\right),
\\
&\boldsymbol{\mu}_t\left(\zv_t, \zv_0\right)=\textstyle\frac{1}{\sqrt{\alpha_t}}\left(\zv_t-\frac{1-\alpha_t}{\sqrt{1-\bar{\alpha}_t}} \boldsymbol{\epsilon}\right),\label{eq:df_rvs_post}
\end{split}
\end{equation}
where $\boldsymbol{\epsilon}$ represents the noise added on $\zv_t$. Thus, a denoising network $\boldsymbol{\epsilon}_\theta$ is used to predict the noise $\boldsymbol{\epsilon}$ at each step, following the previous works~\cite{ho2020denoising,rombach2022high}. In order to encode condition $\yv$ to latent space, another encoder $\text{LE}^\prime$ is applied to extract features, with the same structure as the LE of Phase \uppercase\expandafter{\romannumeral1}. Specifically, $\text{LE}^\prime$ compresses the normalized measurement $\yv_{norm}$ into latent space to get the latent condition features $\cv \in \mathbb{R}^{N \times C}$. In the end, we use the denoising network to predict the noise $\epsilon_t$ according to $\zv_t$ of the previous step in reverse process and the condition $\cv$, stated as $\boldsymbol{\epsilon} = \boldsymbol{\epsilon}_\theta\left(\zv_t, \cv, t\right)$. With the substitution of $\boldsymbol{\epsilon}_\theta$ in Eq.~(\ref{eq:df_rvs_post}) and set the variance as $1 -\alpha_t$, the reverse inference can be stated as:
\begin{equation}
\boldsymbol{z}_{t-1}=\textstyle \frac{1}{\sqrt{\alpha_t}}\left(\boldsymbol{z}_t-\frac{1-\alpha_t}{\sqrt{1-\bar{\alpha}_t}} \boldsymbol{\epsilon}_\theta\left(\zv_t, \cv, t\right)\right)+\sqrt{1-\alpha_t} \boldsymbol{\epsilon}_t, \label{eq:df_rvs_sample}
\end{equation}
where $\boldsymbol{\epsilon}_t \sim \mathcal{N}(0, \mathbf{I})$. Finally, we can generate the target prior feature $\hat{\zv} \in \mathbb{R}^{N \times C}$ after $T$ iterative sampling $\zv_t$ by Eq.~(\ref{eq:df_rvs_sample}). As shown in Fig.~\ref{fig:overall}(e), the predicted prior feature is then used to guide the Transformer in denoiser. Notably, since the distribution of the latent space with the size of $\mathbb{R}^{N \times C}$ (\eg, $16 \times 256$) is much simpler than that of images with size $\mathbb{R}^{H \times W \times N_\lambda}$, the prior feature can be generated with a small number of iterations $T$, corresponding to paper~\cite{rombach2022high}.

Typically, training the diffusion model refers to training the denoising network $\boldsymbol{\epsilon}_\theta$. Following the previous works~\cite{ho2020denoising,song2020denoising}, we train the model by optimizing the weighted variational bound. The training objective is:
\begin{equation}
\nabla_{\boldsymbol{\theta}}\left\|\boldsymbol{\epsilon}-\boldsymbol{\epsilon}_\theta\left(\sqrt{\bar{\alpha}_t} \zv_{GT}+\sqrt{1-\bar{\alpha}_t} \boldsymbol{\epsilon}, \cv, t\right)\right\|_2^2, \label{eq:grad_diff}
\end{equation}
where $\zv_{GT}$ and $\cv$ are ground-truth prior features and the latent condition representations defined above; $t \in[1, T]$ is the randomly sampled time step; $\boldsymbol{\epsilon} \sim \mathcal{N}(0, \mathbf{I})$ denotes the sampled Gaussian noise. 
We employ DDPM~\cite{ho2020denoising} for diffusion as the original LDM~\cite{rombach2022high}. Considering small prior size and time efficiency, we adopt a simple denoising network consisting of several MLP layers for fast diffusion denoising.
All the parameters are jointly updated in the network with the objective loss function of the second phase, including: 

the DUN, the feature encoder $\text{LE}^\prime$, and the diffusion denoising network $\boldsymbol{\epsilon}_\theta$.
The objective loss function of the second phase can be stated as:
\begin{equation}
\mathcal{L}_{\text {diff}}=\|\hat{\zv}-\zv\|_1, \quad
\mathcal{L}_{\text {all}}=\mathcal{L}_{\text {rec}}+\mathcal{L}_{\text {diff}}. \label{eq:loss2}
\end{equation}
Here, we do not use Eq.~(\ref{eq:grad_diff}) as $\mathcal{L}_{\text{diff}}$ because it only trains diffusion at `t'-th step while we execute the all-time-step together and let the LDM directly predict $\hat{\zv}$.
The entire two-phase training procedure is summarized in Algorithm~\ref{algo: training}.

\subsection{Aggregate Features by Trident Transformer}\label{subsec:Trans}
\begin{figure*}
\centering
\includegraphics[width=1\linewidth]{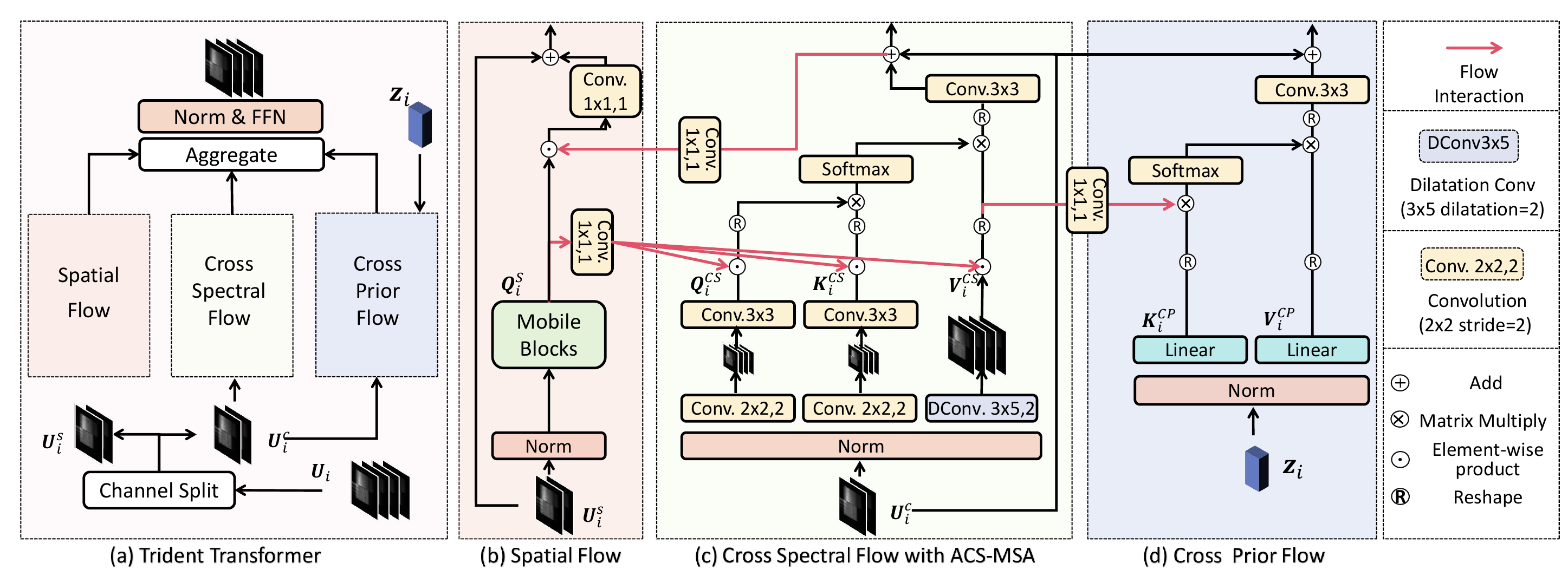}
\caption{\small (a) The Trident Transformer in Fig.~\ref{fig:overall}(d). (b)-(d) are the detailed sub-modules. $\Umat_i$ is the input feature. The prior feature $\Zmat_i$ is sent into the prior flow. }
\label{fig:trans}
\end{figure*}
Previous HSI reconstruction methods usually only exploit the relation between spatial and spectral, both externally and internally. However, the spatial-spectral relations are challenging to explore only with compressed measurements. Therefore,
we design a Transformer, named Trident Transformer (TT), to effectively aggregate high-quality degradation-free prior knowledge for compensation.

Firstly, inspired by the multi-scale operations in previous papers~\cite{chen2023hierarchical,xia2023diffir} with hierarchical structures, we downsample the prior to obtain the multi-scale prior representations along with the U-shape levels in Fig.~\ref{fig:overall}(d). Specifically, three downsampling layers are employed, and the outputs contain prior features of three scales, stated as: 
\begin{equation}
\zv^{i} = \left\{\begin{array}{ccl} 
  \zv_{GT} \text{~or~} \hat{\zv},&{if}&i=1,\\
  \text{downsample}(\zv^{i-1}),&{if}&i>1
\end{array} \right.~,
\end{equation}
where $\zv^{i}\in\mathbb{R}^{\frac{N}{2^i-1} \times 2^{i-1}C}, ~i=1,2,3$. For Phase \uppercase\expandafter{\romannumeral1}, $\zv^{1} = \zv_{GT}$, which is computed in the first phase training; For Phase \uppercase\expandafter{\romannumeral2}, $\zv^{1} = \hat{\zv}$, which is utilized for training and inference in the second phase.   

As shown in Fig.~\ref{fig:trans}, our Trident Transformer includes three branches: spatial flow, cross-spectral flow,  and cross-prior flow. Each branch shares the information flow with others and is then fused by the aggregation layer and a feed-forward network (FFN).
Before the embedding layer, the input feature at $i$-th scale $\Umat_i\in \mathbb{R}^{H_i \times W_i \times C_i}$ is split into  $\Umat^{C}_i\in \mathbb{R}^{H_i \times W_i \times \frac{C_i}{2}}$ and $\Umat^{S}_i\in \mathbb{R}^{H_i \times W_i \times \frac{C_i}{2}}$ along the channel dimension, denoting cross flow input and spatial flow input respectively.
The spatial flow consists of a series of MBlocks without batch norms.

\noindent\textbf{Cross Spectral Flow}
In the cross spectral flow (CSF) module, as shown in Fig.~\ref{fig:trans} (c), we design asymmetric cross-scale multi-head self-attention (ACS-MHSA). Pansharpening (PAN) is a technique of using a high-resolution (HR) panchromatic and a lower-resolution (LR) HSI to generate an HR-HSI. Compared with directly capturing HR-HSI, it requires less amount of data.
Inspired by the PAN, this flow primarily focuses on the spectral dimension and aims to save computational burden according to the spatial size. Specifically, we compress the spatial resolution of the query embedding ($\Qmat$) and key embedding ($\Kmat$) to $\frac{1}{4}$ and expand its channel twice. Considering that CASSI measurements are shifted along one axis, there are more spatial correlations along this axis. Thus, after establishing the spectral correlation, we use an asymmetric dilation convolution (DConv) with kernel size $3\times5$ on the value embedding ($\Vmat$) to obtain larger perceptual field information along the shifted axis with expanded channel dimension and unchanged spatial dimension. Embedding $\Qmat$, $\Kmat$, $\Vmat$, and spatial compensation $\Pmat_{i}^{S}$ from spatial flow are formulated as: 
\begin{align}
\Qmat_{i}^{CS} &= \Wmat^{QCS}\Umat_{i},&
\Kmat_{i}^{CS} &= \Wmat^{KCS}\Umat_{i},&
\Vmat_{i}^{CS} &= \Wmat^{VCS}\Umat_{i},   \\
\Pmat_{i}^{SQK} &= \downarrow(\Wmat^{PSQ}\Qmat_{i}^{S}),&
\Pmat_{i}^{SV} &= \Wmat^{PSV}\Qmat_{i}^{S},& 
\Mmat_i &= \Qmat_{i}^{CS}(\Kmat_{i}^{CS})^{\top},
\end{align}
where $\Wmat^*$ represents the weights of bias-free convolution, and $\downarrow$ is downsampling. The procedure of asymmetric cross-scale self-attention can be formulated as:
\begin{align}
\text{ACS-MHSA}_{i}(\Umat_{i}) =\Pmat_{i}^{SV}\odot\Wmat_{c1}^{CS}\Vmat_{i}^{CS}\text{Softmax}(\Pmat_{i}^{SQK}\odot\Mmat_i/\alpha), \label{eq:sec4-attention_spec} 
\end{align}
\noindent\textbf{Cross Prior Flow}
Cross prior flow (CPF) in Fig.~\ref{fig:trans} (d) is a variable shared multi-head cross-attention. The query in this flow is borrowed from the value of CSF which is extracted from a large perceptive field with more spatial information. In this way, the prior could facilitate to compensate for spatial deficiency. Compared to the spectral recovery, the spatial recovery is typically more challenging. Our manipulation can be formulated as:
\begin{align}
& {MHSA_{i}^{CP}}(\Umat_{i}) =W_{c1}\Vmat\odot \text{Softmax}(\Kmat\odot\Qmat/\alpha), \label{eq:sec4-attention_prior} \\
& \Qmat_{i}^{CP} = \Qmat_{i}^{CS},~~\Kmat_{i}^{CP} = \Wmat^{K}_{z}\zv_{i},~~\Vmat_{i}^{CP} = \Wmat^{V}_{z}\zv_{i},  
\end{align}
where $ \zv^{i}, i=1,2,3$ denotes the prior feature of different spatial levels.

\noindent\textbf{Flow Interaction and Aggregation}
In order to compensate for the deficiency of spatial information in CSF and CPF, and the spectral information in the spatial flow, we fuse the compensation information together to reconstruct hyperspectra images. 
As shown in Fig.~\ref{fig:trans}, the colorful arrows represent the information interactions between each flow. Specifically, information of each module modulates with other flows, where the $1\times 1$ convolutions serve as compensation bridges. 
The aggregation part consists of concatenation, convolution layers, and an activation function (details can be seen in SM) for a weighted combination of each flow output.
In our Trident Transformer, the prior knowledge learned from the clean images will provide compensation for reconstruction in both spatial and spectral details, avoiding the influence of degraded measurements.

\begin{center} 
\begin{minipage}{0.87\linewidth}
\begin{algorithm}[H]
        \caption{\small Two-phase Training Strategy} \label{algo: training}
	\begin{algorithmic}[1]
        \Require
        Dataset $\mathcal{D}=\{(\xv^{(n)},\yv^{(n)}\}^N_{n=1}$; Sensing matrix $\Av$;
        Random initialized parameter of DUN $\phi_{DUN}$, latent encoder LE network $\phi_{LE}$, conditional encoder LE' network $\phi_{CLE}$, diffusion denoising network $\phi_{\epsilon}$; 
\While {Not Converge} \textcolor{gray}{\Comment{Phase \uppercase\expandafter{\romannumeral1} training,~`$\leftarrow$' denotes update}} 
                 \State $\zv_{GT} \leftarrow \text{LE}(\Iv_{\text{E}}|\phi_{LE})$; $\hat{x} \leftarrow \text{DUN}(\yv,\Av,\zv_{GT}|\phi_{DUN})$ 
                 
                 \State Jointly update $\phi_{DUN} \text{ and } \phi_{LE}$ by $\mathcal{L}_{rec}$;
        
\EndWhile
        \State Freeze $\phi_{LE}$;
		\While {Not Converge} \textcolor{gray}{\Comment{Phase \uppercase\expandafter{\romannumeral2} training}}
                 \State $\cv \leftarrow \text{LE}^\prime(\yv_{norm}|\phi_{CLE})$;
                 $\hat{\zv} \leftarrow \text{Diff}(\cv|\phi_{\epsilon})$; 
                 
                 \State $\zv_{GT} \leftarrow \text{LE}(\Iv_{\text{E}}|\phi_{LE})$;    
                 $\hat{x}  \leftarrow \text{DUN}(\yv,\Av,\hat{\zv})|\phi_{DUN})$;
                 
                 \State Jointly update $\phi_{DUN},\phi_{CLE}, \text{ and } \phi_{\epsilon}$ by $\mathcal{L}_{all}$ in Eq.~(\ref{eq:loss2});
        \EndWhile
	\end{algorithmic}
\end{algorithm}
\end{minipage}
\end{center}
\begin{figure*}
\centering
\includegraphics[width=1\linewidth]{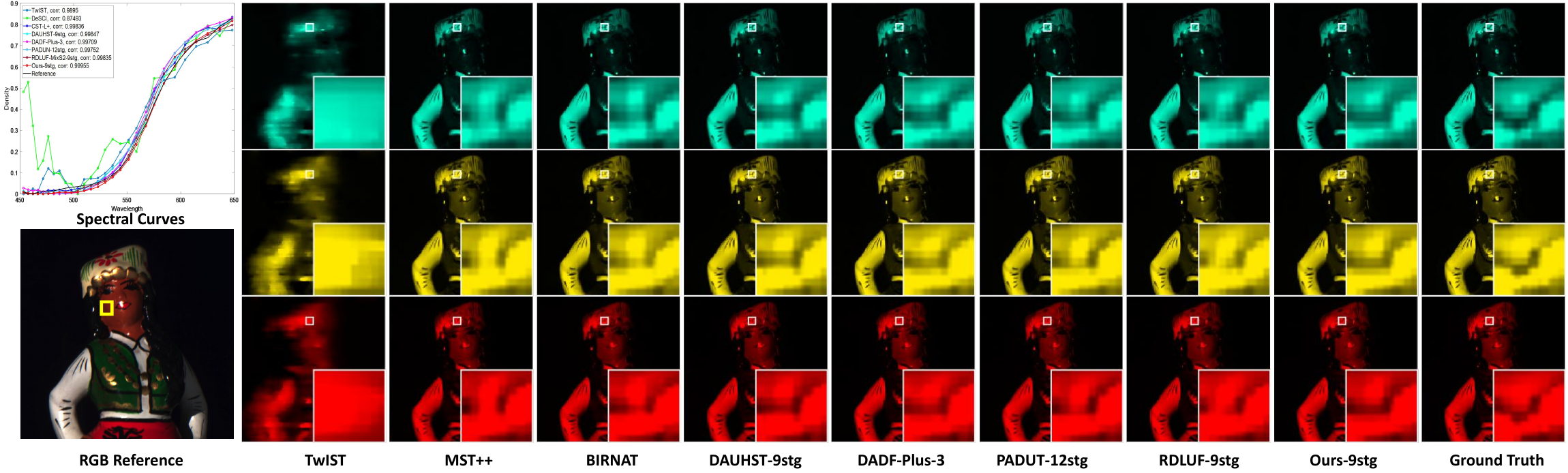}
\caption{\small The visualization result on synthetic data. 3 out of 28 wavelengths are selected for visual comparison. `Corr' in the top left curve is the correlation coefficient between one method curve and the ground truth curve of the chosen (golden box) region.}
\label{fig:sim_fig}
\end{figure*}
\begin{table*}
  \caption{The results of PSNR in dB (top entry in each cell), SSIM (bottom entry in each cell) on the 10 synthetic spectral scenes.`-3stg' denotes the network with 3 unfolding stages. `Avg'  represents the average of 10 scenes. \textbf{Bold}: Best.}
\begin{center}
  \resizebox{\textwidth}{!}{
  \begin{tabular}{c|cccccccccc|c}
    \toprule
Algorithms & Scene1 & Scene2 & Scene3 & Scene4 & Scene5 & Scene6 & Scene7 & Scene8 & Scene9 & Scene10 & Avg \\ \midrule
\multirow{2}{*}{TwIST}                  & 25.16                 & 23.02                 & 21.40                 & 30.19                 & 21.41                 & 20.95                 & 22.20                 & 21.82                 & 22.42                 & 22.67                 & 23.12                  \\
                       & 0.700                 & 0.604                 & 0.711                 & 0.851                 & 0.635                 & 0.644                 & 0.643                 & 0.650                 & 0.690                 & 0.569                 & 0.669                  \\ 
\midrule
\multirow{2}{*}{DNU}                    & 31.72                 & 31.13                 & 29.99                 & 35.34                 & 29.03                 & 30.87                 & 28.99                 & 30.13                 & 31.03                 & 29.14                 & 30.74                  \\
                       & 0.863                 & 0.846                 & 0.845                 & 0.908                 & 0.833                 & 0.887                 & 0.839                 & 0.885                 & 0.876                 & 0.849                 & 0.863                  \\ 
\midrule
\multirow{2}{*}{MST++}                  & 35.40                 & 35.87                 & 36.51                 & 42.27                 & 32.77                 & 34.80                 & 33.66                 & 32.67                 & 35.39                 & 32.50                 & 35.99                  \\
                       & 0.941                 & 0.944                 & 0.953                 & 0.973                 & 0.947                 & 0.955                 & 0.925                 & 0.948                 & 0.949                 & 0.941                 & 0.951                  \\ 
\midrule
\multirow{2}{*}{BIRNAT}                  & 36.79 & 37.89 & 40.61 & 46.94 & 35.42 & 35.30 & 36.58 & 33.96 & 39.47 & 32.80 & 37.58  \\
& 0.951 & 0.957 & 0.971 & 0.985 & 0.964 & 0.959 & 0.955 & 0.956 & 0.970 & 0.938 & 0.960                   \\ 
\midrule
\multirow{2}{*}{LRSDN} & 35.44 & 34.89 & 38.90 & 45.29 & 34.71 & 33.18 & 37.76 & 30.57 & 39.49 & 30.62 & 36.08 \\
      & 0.923 & 0.909 & 0.961 & 0.985 & 0.949 & 0.930 & 0.964 & 0.901 & 0.963 & 0.889 & 0.938 \\
   \midrule   
\multirow{2}{*}{DAUHST-9stg}            & 37.25                 & 39.02                 & 41.05                 & 46.15                 & 35.80                 & 37.08                 & 37.57                 & 35.10                 & 40.02                 & 34.59                 & 38.36                  \\
                       & 0.958                 & 0.967                 & 0.971                 & 0.983                 & 0.969                 & 0.970                 & 0.963                 & 0.966                 & 0.970                 & 0.956                 & 0.967                  \\ 
\midrule
\multirow{2}{*}{DADF-Plus-3}              
& 37.46                 & 39.86                 & 41.03                 & 45.98                 & 35.53                 & 37.02                 & 36.76                 & 34.78                 & 40.07                 & 34.39                 & 38.29                  \\
&0.965 &0.976 &0.974 &0.989 &0.972 &0.975 &0.958 &0.971 &0.976 &0.962                 & 0.972                  \\ 
\midrule
\multirow{2}{*}{PADUT-5stg}      & 36.68 &38.74 &41.37 &45.79 &35.13 &36.37 &36.52 &34.40 &39.57 &33.78 &37.84                  \\
                       & 0.955 &0.969 &0.975 &0.988 &0.967 &0.969 &0.959 &0.967 &0.971 &0.955 &0.967                  \\ 
\midrule
\multirow{2}{*}{RDLUF-MixS2-3stg}       & 36.67                 & 38.48                 & 40.63                 & 46.04                 & 34.63                 & 36.18                 & 35.85                 & 34.37                 & 38.98                 & 33.73                 & 37.56                  \\
                       & 0.953                 & 0.965                 & 0.971                 & 0.986                 & 0.963                 & 0.966                 & 0.951                 & 0.963                 & 0.966                 & 0.950                 & 0.963                  \\ 
\midrule
\rowcolor[rgb]{0.898,0.898,0.902} {\cellcolor[rgb]{0.898,0.898,0.902}}              & 37.14&  39.60&  41.78&  46.57&  35.57&  37.02&  36.80&  35.22&  40.15&  34.17& 38.40\\
\rowcolor[rgb]{0.898,0.898,0.902} \multirow{-2}{*}{{\cellcolor[rgb]{0.898,0.898,0.902}}Ours-3stg}                      & 0.963&  0.975&  0.978&  0.990&  0.971&  0.975&  0.960&  0.973&  0.976&  0.962& 0.972                  \\ 
\midrule
\rowcolor[rgb]{0.898,0.898,0.902} {\cellcolor[rgb]{0.898,0.898,0.902}}             & 37.88&  40.92&  43.41&  47.18&  37.12&  37.74&  38.28&  35.73&  41.48&  35.18&  39.38\\
\rowcolor[rgb]{0.898,0.898,0.902} \multirow{-2}{*}{{\cellcolor[rgb]{0.898,0.898,0.902}}Ours-5stg}                        & 0.968&  0.980&  0.983&  0.992&  0.978&  0.980&  0.969&  0.977&  0.981&  0.967&   0.977                  \\ 
\midrule
\multirow{2}{*}{PADUT-12stg}      & 37.36 &40.43 &42.38 &46.62 &36.26 &37.27 &37.83 &35.33 &40.86 &34.55 &38.89                 \\
                       & 0.962 &0.978 &0.979 &0.990 &0.974 &0.974 &0.966 &0.974 &0.978 &0.963 &0.974                  \\ 
\midrule
\multirow{2}{*}{RDLUF-MixS2-9stg}                             & 37.94          & 40.95          & 43.25          & 47.83          & 37.11          & 37.47          & 38.58          & 35.50          & 41.83          & 35.23          & 39.57           \\
                                             & 0.966          & 0.977          & 0.979          & 0.990          & 0.976          & 0.975          & 0.969          & 0.970          & 0.978          & 0.962          & 0.974           \\ 
\midrule
\rowcolor[rgb]{0.898,0.898,0.902} {\cellcolor[rgb]{0.898,0.898,0.902}}   & \textbf{38.08} & 41.84          & 43.77          & 47.99          & 37.97          & 38.30          & 38.82          & 36.15          & 42.53          & 35.48          & 40.09                    \\
\rowcolor[rgb]{0.898,0.898,0.902} \multirow{-2}{*}{{\cellcolor[rgb]{0.898,0.898,0.902}}Ours-9stg}           & 0.969          & 0.982          & 0.983          & \textbf{0.993} & 0.980          & 0.980          & 0.973          & \textbf{0.979} & \textbf{0.984} & \textbf{0.970} & 0.979                    \\ 
\midrule
\rowcolor[rgb]{0.898,0.898,0.902} {\cellcolor[rgb]{0.898,0.898,0.902}}   & \textbf{38.08} & \textbf{41.85} & \textbf{43.83} & \textbf{48.04} & \textbf{38.00} & \textbf{38.32} & \textbf{38.94} & \textbf{36.20} & \textbf{42.81} & \textbf{35.54} & \textbf{40.16}           \\
\rowcolor[rgb]{0.898,0.898,0.902} \multirow{-2}{*}{{\cellcolor[rgb]{0.898,0.898,0.902}}Ours-10stg}           & \textbf{0.970} & \textbf{0.984} & \textbf{0.984} & \textbf{0.993} & \textbf{0.982} & \textbf{0.982} & \textbf{0.974} & \textbf{0.979} & \textbf{0.984} & \textbf{0.970} & \textbf{\textbf{0.980}}  \\
\bottomrule
  \end{tabular}}
  \end{center}
  \label{Table:sim-grayscale}
\end{table*}
\begin{table}
\centering
\caption{ \small Performance and computational efficiency comparisons with recent methods. 
}
\resizebox{1\linewidth}{!}{
\begin{tabular}{c|c|c|c|c|c|c} 
\hline
\hline
 Method   & Venue  & PSNR~(dB) & Params~(M)  & FLOPs~(G) & Infer. Time~(ms) & Training Time~(h)\\ 
\hline
PADUT-12stg & ICCV'23 & 38.89 & 5.38 & 90.46  & \textbf{749.94}   & \textbf{123.3}\\
RDLUF-9stg  & CVPR'23 & 39.57 & \textbf{1.89} & 231.09 & 913.34    & 155.5\\
\rowcolor[rgb]{0.898,0.898,0.902} Ours-9stg & - & \textbf{40.09} & 2.78 & \textbf{88.68}   & 1096.58  & 143.6\\
\hline
\hline
\end{tabular}}\label{tab:time_param_flops}
\end{table}

\begin{table}
\centering
\caption{Ablation study of different modules in our 3-stage unfolding network. `PSNR' is the average of 10 synthetic scenes. `FLOPs~(G)' denotes the FLOPs in testing.}
\resizebox{1\linewidth}{!}{
\begin{tabular}{c|c|c|c|c|c|c|c|c}
\hline
\hline
Method             & Projection                      & SF                          & CSF Atten.                         & CPF                        & LE Input                  & Diffusion                   & PSNR  & FLOPs (G)  \\
\hline
w/o GC-GAP         & Basic GAP & \Checkmark   & ACS   & \Checkmark   & $\Iv_{\text{E}}$ & \Checkmark   & 38.01 & 29.84      \\
w/o spatial flow   & GC-GAP   & \XSolidBrush & ACS   & \Checkmark   & $\Iv_{\text{E}}$ & \Checkmark   & 37.66 & 30.13       \\
Basic MHSA  & GC-GAP   & \Checkmark   & Basic & \Checkmark   & $\Iv_{\text{E}}$ & \Checkmark   & 38.31 & 36.80      \\

w/o prior flow     & GC-GAP   & \Checkmark   & ACS   & \XSolidBrush & $\Iv_{\text{E}}$ & \Checkmark   & 37.89 & 30.37      \\
Inaccurate guidance & GC-GAP   & \Checkmark   & ACS   & \Checkmark   &  $\yv$       & \Checkmark   & 37.63 & 33.80      \\
w/o diffusion      & GC-GAP  & \Checkmark   & ACS   & \Checkmark   & $\Iv_{\text{E}}$ & \XSolidBrush & 37.61 & 33.70      \\
\rowcolor[rgb]{0.898,0.898,0.902} Our full model     & GC-GAP   & \Checkmark   & ACS   & \Checkmark   & $\Iv_{\text{E}}$ & \Checkmark   & 38.40 & 33.80   \\  
\hline
\hline
\end{tabular}}
\label{Tab:Abla_modules}
\end{table}
\begin{figure*} 
\centering
\includegraphics[width=1\linewidth,trim=0 1 0 0,clip]{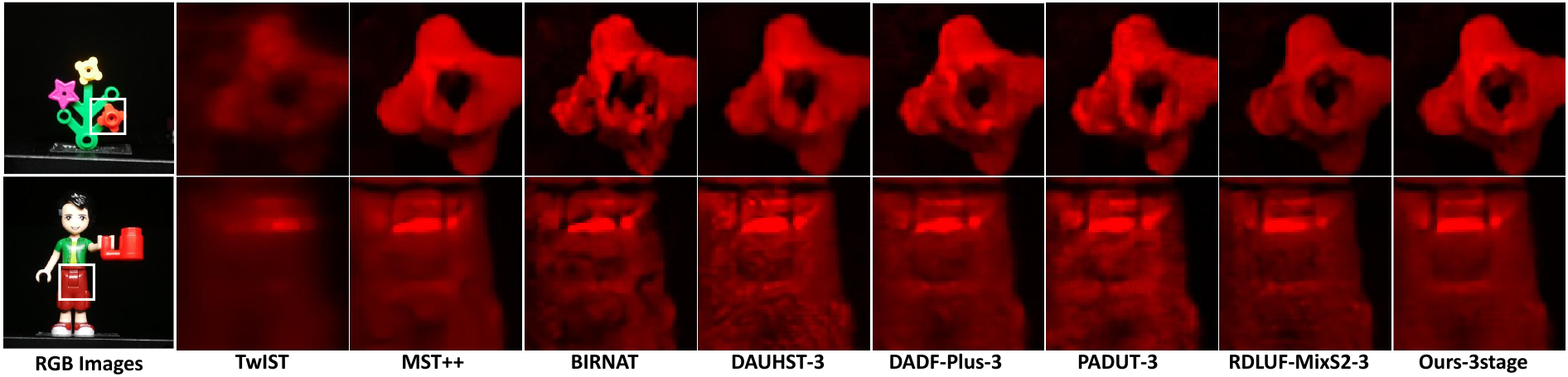}
\caption{\small The real data visual comparisons. 2 out of 28 wavelengths are selected.}
\label{fig:real_fig}
\end{figure*}
\section{Experiments}
We conduct experiments on both simulation and real HSI datasets. 
Following the approaches in~\cite{Meng20ECCV_TSAnet,meng2020gap,huang2021deep,cai2022mask}, we select a set of 28 wavelengths ranging from 450-650nm by employing spectral interpolation applied to the HSI data.
\subsection{Experimental Settings}
\textbf{Simulation and Real Datasets:} We adopt two widely used HSI datasets, i.e., CAVE~\cite{park2007multispectral} and KAIST~\cite{choi2017high} for simulation experiments. The CAVE dataset comprises 32 HSIs with a spatial size of 512 $\times$ 512. The KAIST dataset includes 30 HSIs with a spatial size of 2704 $\times$ 3376. Following previous works~\cite{Meng20ECCV_TSAnet,meng2020gap,huang2021deep,cai2022mask}, we only employ the CAVE dataset as the training set both in Phase \uppercase\expandafter{\romannumeral1} and \uppercase\expandafter{\romannumeral2}, while 10 scenes from the KAIST dataset are utilized for testing. During the training process, a real mask of size 256 $\times$ 256 pixels is applied.
In our real experiment, we utilized the HSI dataset captured by the SD-CASSI system in~\cite{Meng20ECCV_TSAnet}. The system captures real-world scenes of size 660 $\times$ 714$ \times$ 28 with wavelengths spanning from 450 to 650 nm and dispersion of 54 pixels.

\textbf{Implementation Details:} 
For the diffusion settings, the iteration number $T$ of the diffusion is set to 16, and the latent space dimension $N$ is set to 16. 
Training Phase \uppercase\expandafter{\romannumeral1} (train DUN and LE) needs 200 epochs and Phase \uppercase\expandafter{\romannumeral2} (train DU, $\boldsymbol{\epsilon}_\theta$, and LE') needs 100 epochs. For all phases of training, we use the Adam~\cite{kingma2014adam} optimizer and cosine scheduler.
PSNR and SSIM~\cite{wang2004image} are utilized as our metrics.  Our method is implemented with the
PyTorch and trained using NVIDIA RTX3090 GPUs. More details can be seen in the supplementary material (SM).
\subsection{Compare with State-of-the-art}
We compare our method with previous methods including the end-to-end networks: DADF-Net~\cite{xu2023degradation}, MST~\cite{cai2022mask}, BIRNAT~\cite{cheng2022recurrent}; the deep unfolding methods: RDLUF-MixS2~\cite{dong2023residual}, PADUT~\cite{li2023pixel}, DAUHST~\cite{cai2022degradation}, DNU~\cite{wang2020dnu}; the self-supervised method: LRSDN~\cite{chen2024hyperspectral}; and traditional model-based method: TwIST~\cite{Bioucas-Dias2007TwIST}. The comparisons are conducted on both synthetic and real datasets.

\textbf{Synthetic data:} 
The numeric comparisons on synthetic data can be seen in Table~\ref{Table:sim-grayscale}. Our proposed method surpasses the recent SOTA method RDLUF-MixS2 according to average PSNR (+0.59 dB) and SSIM (+0.06).
Fig.~\ref{fig:sim_fig} shows the visual reconstruction results. Three wavelengths including striking colors in RGB reference red, yellow, and green are selected to compare. The golden box part in the reference was chosen to calculate and compare the wavelength accuracy. The accuracy metric is the correlation coefficient with the ground truth of the chosen region, \ie the `Corr’ in the curves. 
According to the `Corr’, our method (0.9995) has a more accurate wavelength curve than others. The zoomed part in the figure also demonstrates that our method has clearer edges on the hat than others.
In Table.~\ref{tab:time_param_flops}, we list the performance, parameter number, FLOPs, inference time, and training time of our method and other recent unfolding methods. `Infer. Time’ denotes the total inference time of each method dealing with 10 synthetic test scenes. `Training Time' is under 300 total epochs setting.  
The table illustrates that our methods exhibit superior performance, the lowest FLOPs, along with reasonable parameter counts and times. 
\textbf{Real data:} 
Two scenes of real SD-CASSI measurement reconstruction results are shown in Fig.~\ref{fig:real_fig}, and two obvious color regions in RGB references are selected to compare. Our method shows less noise and artefacts on the plastic toy surface.
\section{Ablation Study}
In this ablation study, we train our model on the synthetic training data with models with 3 unfolding stages. The results are summarized in Table~\ref{Tab:Abla_modules}. `Inaccurate guidance' denotes that we only use measurement as the latent encoder input instead of clean HSIs.  `w/o prior flow' denotes that using simple MLP layers instead of CPF. `w/o diffusion' denotes removing diffusion in Phase \uppercase\expandafter{\romannumeral2}.
\begin{tabwindow}[0,r,
  {\resizebox{0.34\textwidth}{!}{
\begin{tabular}{c|c|c|c|c} 
    \hline
    \hline
    $\zv$ scale ($N$) & 4 & \cellcolor{color3}16 & 64 & 256                           \\ 
    \hline
    PSNR & 37.99 & \cellcolor{color3}38.40 & 38.40 & 38.41      \\ 
    FLOPs & 33.31 & \cellcolor{color3}33.80 & 34.49 & 36.99                                   \\
    \hline
    \hline
\end{tabular} }
},{$\zv$ scale comparisons.\label{tab:abla_z_scale}}]
 The ablation illustrates that with LDM prior assistance, we can achieve better reconstruction results, and our design of the Trident Transformer successfully aggregates three types of information and effectively compensates for some reconstruction defects. Fig.~\ref{fig:teaser_vis} also visualizes the feature map changes before and after prior enhancement. The enhanced features demonstrate increased concentration on edges and reduced noise.
Moreover, without accurate guidance, the LDM will even harm the reconstruction. The method `Basic MHSA' illustrates that ACS-MHSA in CSF has better performance and higher computational efficiency than basic MHSA.
The diffusion steps shown in Fig.~\ref{fig:teaser_bubble}(b) illustrate that 16 steps are enough for good reconstruction results. Table~\ref{tab:time_param_flops} illustrates that inference time is still in a reasonable range even using diffusion 16 steps.
For the scale of $\zv$, we compared different values of $N$ in Table~\ref{tab:abla_z_scale}, and we find that $N=16$ can keep the balance of performance and efficiency. More ablation studies can be seen in SM.\end{tabwindow}
\section{Conclusion}
In this paper, we introduce a novel deep unfolding network that leverages prior knowledge from the latent diffusion model for spectral reconstruction.
It achieves SOTA performance on both simulated data and real data.

\section*{Acknowledgements}
This work was supported by the National Natural Science Foundation of China (grant number 62271414), Zhejiang Provincial Outstanding Youth Science Foundation (grant number LR23F010001), Zhejiang ``Pioneer'' and ``Leading Goose'' R\&D Program  (grant number 2024SDXHDX0006, 2024C03182), the Key Project of Westlake Institute for Optoelectronics (grant number 2023GD007), and the 2023 International Sci-tech Cooperation Projects under the purview of the ``Innovation Yongjiang 2035'' Key R\&D Program  (grant number 2024Z126).

{
    \bibliographystyle{splncs04}
    \bibliography{string,main}
}

\end{document}